# Atomic-scale Control of Tunnel Coupling

Xiqiao Wang[1,2,3 #], Jonathan Wyrick[1 #], Ranjit V. Kashid[1], Pradeep Namboodiri[1], Scott W. Schmucker[3], Andrew Murphy[1 ^], M. D. Stewart, Jr.[1], Neil Zimmerman[1], and Richard M. Silver[1*]

**Atomically precise silicon-phosphorus (Si:P) quantum systems are actively being pursued to realize universal quantum computation[1] and analog quantum simulation.[2] Atomically precise control of the tunneling coupling strength is critical to spin-exchange operations in tunnel-coupled quantun dots and spin-selective tunneling for initialization and read-out .[3, 4, 5] It was shown by Pok[6] and Fuhrer *et al*.[7] that engineering the nanogap geometry, can affect the tunnel barriers and the tunnel rates in STM-patterned devices: even a ~1 nm difference in the tunnel gap distance can drastically change the tunnel coupling and transport properties in atomically precise Si:P devices.[8] However, critical challenges in atomically precise fabrication have meant systematic, atomic scale control of the tunneling coupling has not been demonstrated. Here using a room-temperature grown locking layer and precise control over the atomic-scale fabrication process, we demonstrate atomic-scale control of the tunnel coupling in atomically precise single electron transistors (SETs). Using the naturally occurring Si (100) 2×1 surface reconstruction lattice as an atomically-precise ruler, we systematically vary the number of lattice counts within the tunnel junction gaps and demonstrate exponential scaling of the tunneling resistance at the atomic limit. We**

---
[1] *National Institute of Standards and Technology, 100 Bureau Dr., Gaithersburg, Maryland 20899, USA*

[2] *Chemical Physics Program, University of Maryland, College Park, Maryland 20742, USA*

[3] *Joint Quantum Institute, University of Maryland, College Park, Maryland 20742, USA*

[#] These authors contributed equally to this work

[^] Current address: *Johns Hopkins University Applied Physics Laboratory, Laurel, Maryland 20723, USA*

**characterize the tunnel coupling asymmetry in a pair of nominally identical tunnel gaps and show a resistance difference of four that corresponds to half a dimer row difference in the effective tunnel gap - the intrinsic limit of hydrogen lithography precision on Si (100) 2×1 surfaces. Our results demonstrate the key capability to do atom-scale design and engineering of the tunnel coupling.**

In this study, we overcome previous challenges by uniquely combining atomically abrupt hydrogen lithography [9, 10] with recent progress in low-temperature epitaxial overgrowth using a locking-layer technique [11, 12, 13] and silicide electrical contact formation [14] to substantially reduce unintentional dopant movement. These advances have allowed us to demonstrate the exponential scaling of the tunneling resistance on the tunnel gap distance at the atomic limit in a systematic and reproducible manner. We define the tunnel gaps with atomically abrupt edges using ultra-clean hydrogen lithography while utilizing the Si (100) surface reconstruction lattice to quantify the tunnel gap distances with atomic accuracy. We suppress unintential dopant movement at the atomic scale using an optimized, room-temperature grown locking layer, which not only locks the dopant position within lithographically defined regions during encapculation overgrowth but also improves device reproducibility by minimizing the impact of overgrowth temperature variations on dopant confinement profiles.[11] Furthermore, our recent development of a high-yield, low-temperature method for forming ohmic contact to burried atomic devices enables robust electrcal characteriation of STM-patterned devices with minimum thermal impact on dopant confinement.[14] With improved capabilities to define and maintain atomically abrupt dopant confinement in silicon, we fabricated a series of STM-patterned Si:P single electron transistors (SETs), where we systematically vary the tunnel junction gap distance with atomic precision, and have used them to demonstrate and explore atomic-scale control of the tunnel coupling. Instead of geometrically simpler single tunnel junctions, we chose SETs in this study because observation of the Coulomb blockade signature is a direct indication that conductance is through the STM-patterned tunnel junctions.

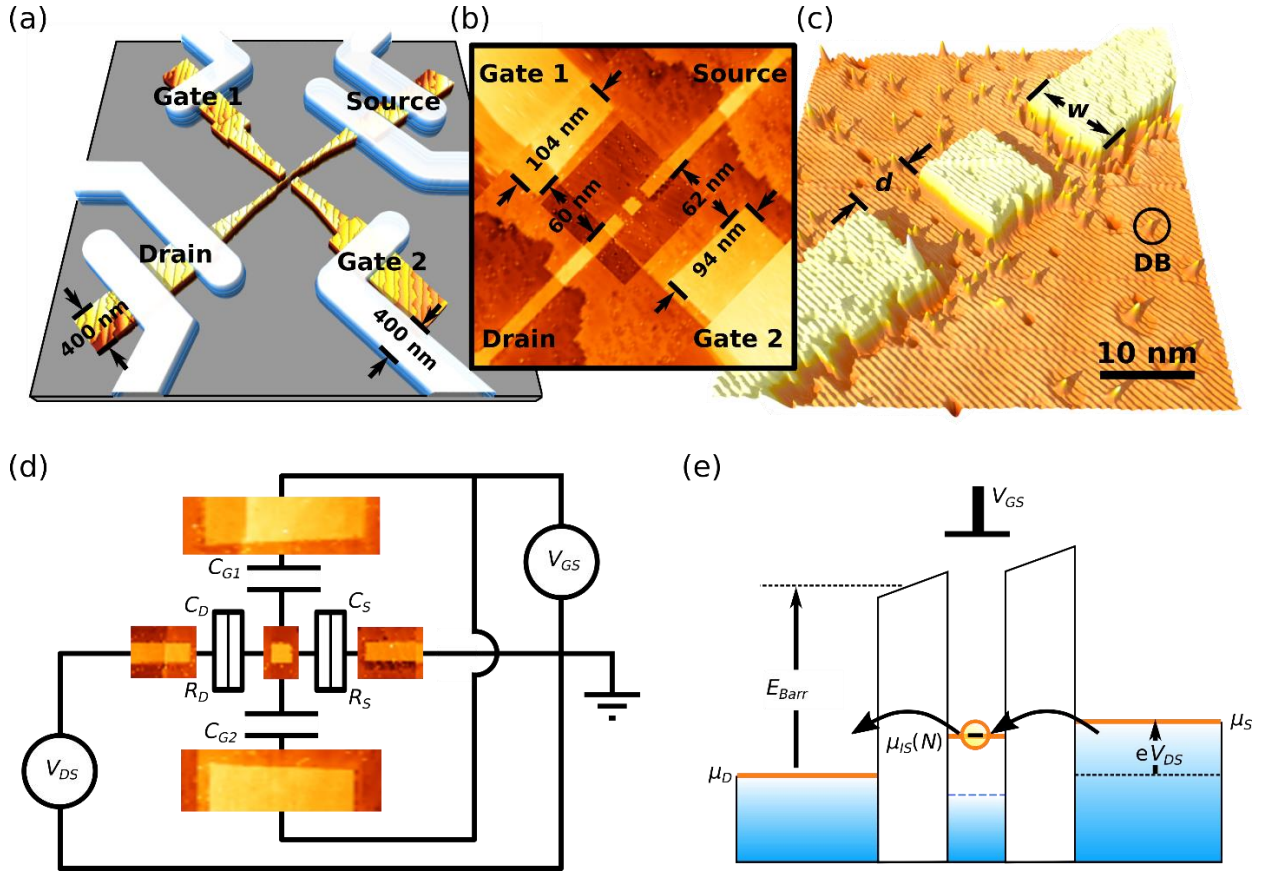

Figure 1. (a) Electrical contacts (sketched in white) overlaid on top of an STM-patterned SET device. (b) STM image of the central device region of a typical SET device acquired immediately following hydrogen lithography. The bright areas are STM-patterns where the hydrogen-resist has been removed, exposing the chemically reactive dangling bonds. The central device region shows a central island that is tunnel coupled with source and drain leads and capacitively coupled to two in-plane gates. Gate 2 is patterned with a deliberate shift towards the source electrode to allow tuning the tunnel coupling symmetry. A high-resolution STM image at the center region is overlaid on a large-scale lower-resolution STM image. (c) Atomic resolution STM image of an SET pattern (SET-C in Figure 2) where the tunnel gaps are defined with atomic precision. The imaged rows running from upper left to lower right are $2 \times 1$ surface reconstruction dimer rows on the Si (100) surface. The junction gap distance, $d$, and junction width, $w$, are marked in the image. The circle marks the image of a single dangling bond (DB).

The STM image is taken at -2 V sample bias and 0.1 nA setpoint current. (d) An equivalent circuit diagram for the SET, where tunnel junctions are treated as a tunneling resistance and capacitance connected in parallel and the combined coupling of the two gates to the SET island is treated as a capacitor. (e) The energy diagram of an SET, where $\mu_S$ and $\mu_D$ are the chemical potentials of the source and drain leads respectively; $\mu_{IS}(N)$ is the chemical potential of the island that is occupied with $N$ excess electrons. $E_{Barr}$ is the barrier height defined by the energy difference between the electrodes' Fermi levels and the conduction band edge of the substrate. We assume a rectangular barrier shape in this study.

Figure 1 (b) shows an STM image of the atomically precise central region of a typical SET device after hydrogen-lithography, but before phosphine dosing. P dopants only incorporate into the bright regions where the STM tip has removed H surface termination atoms and exposed chemically reactive Si-dangling bonds. The Si (100) 2x1 surface reconstruction features dimer rows of pitch $0.77\ nm$ that can serve as a natural "atomic ruler" allowing us to define the critical dimensions with atomic precision. (Figure 1 (c)) The planar source and drain, island (quantum dot), and gates are saturation-dosed resulting in degenerate dopant densities over three orders of magnitude beyond the Mott metal-insulator transition.[15] The island is capacitively coupled to the two in-plane gates through an effective capacitance $C_G$ and to the source (drain) electrodes through tunnel barriers represented by a tunneling resistance $R_S$ ($R_D$) and a capacitance $C_S$ ($C_D$), where each resistance is coupled in parallel with its respective capacitance (Figure 1 (d)). The gate voltages applied to both gates tune the local electrochemical potential of the island and modulate the source-drain current flowing through the central island. Single electrons tunnel sequentially through both barriers due to the electron addition energy (charging effect) on the island.[16] (Figure 1 (e))

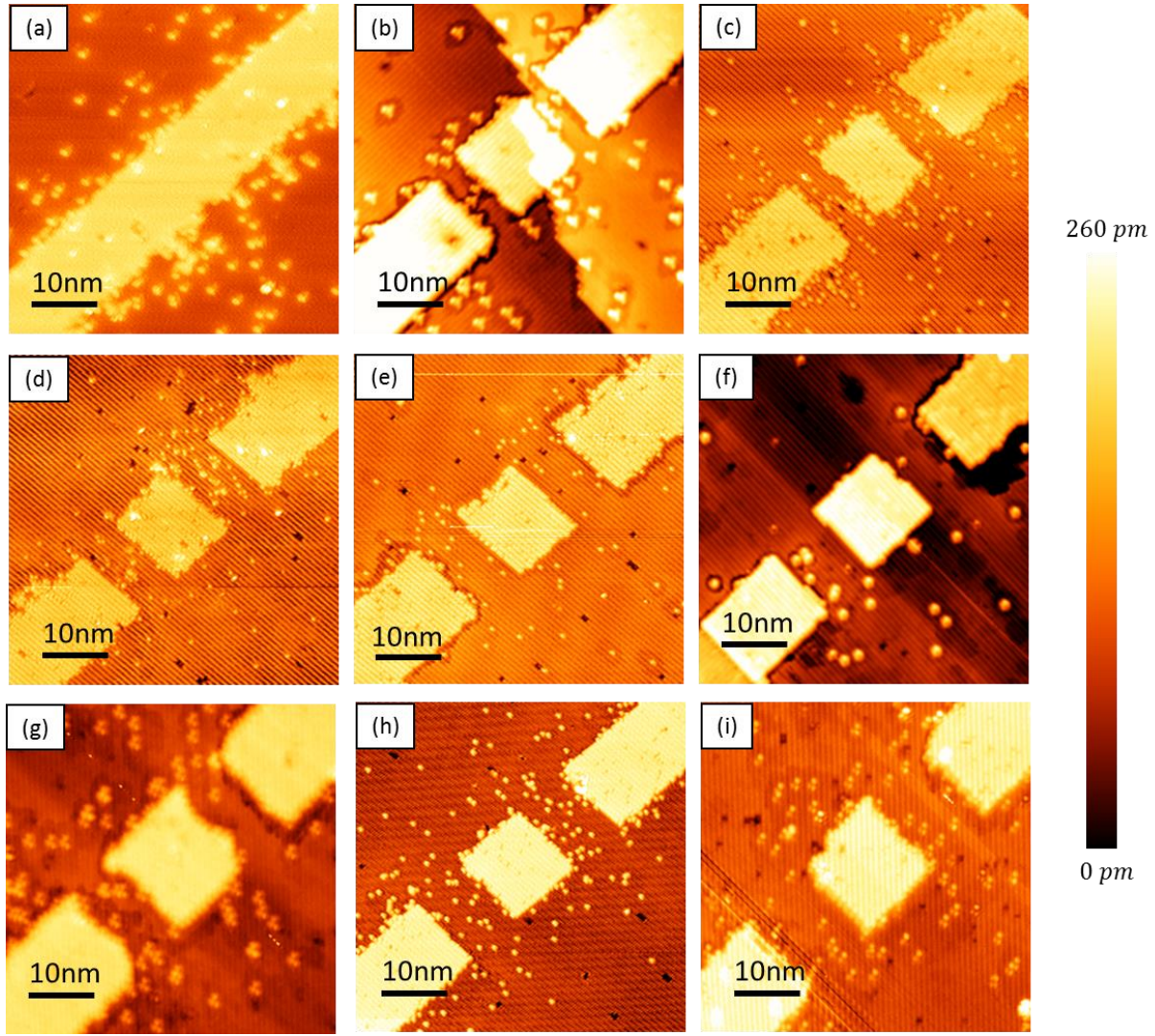

Figure 2. High-resolution STM topography images of the hydrogen-lithography patterns of the wire and SET devices in this study. The devices are named as Wire-A, and SET-B to SET-I according to the figure labels. The drain/source electrodes are oriented in the [110] lattice direction for all cases except for SET-G and SET-I whose drain/source electrodes are oriented in the [100] lattice direction (45° to the [110] direction). Different STM tips/tip conditions are used for the STM images under imaging conditions: -2 V sample bias and 0.1 or 0.05 nA setpoint current.

| Device | Gap distance $d$ (dimer rows) | Lead/island width $w$ (dimer rows) | Island length (dimer rows) | # of squares in leads | $R_S + R_D$ ($M\Omega$) |
|---|---|---|---|---|---|
| Wire-A | 0 | 15.5±1.4 | N/A | 57±4 | N/A |
| SET-B | 7.4±0.6 | 15.3±0.6 | 15.2±0.8 | 74±6 | 0.011±0.009 |
| SET-C | 9.5±0.7 | 15.7±0.7 | 13.1±0.6 | 56±4 | 0.113±0.061 |
| SET-D | 11.1±0.7 | 15.0±0.8 | 14.3±0.6 | 52±4 | 0.340±0.101 |
| SET-E | 11.7±0.4 | 17.5±1.0 | 14.9±0.5 | 56±4 | 2.06±0.69 |
| SET-F | 11.8±0.6 | 15.2±0.4 | 15.3±0.4 | 62±5 | 2.49±0.63 |
| SET-G | 12.2±1.4 | 18.8±1.2 | 17.0±1.5 | 49±4 | 5.55±2.91 |
| SET-H | 13.5±0.6 | 15.1±0.3 | 15.4±0.7 | 52±4 | 127±59 |
| SET-I | 16.2±0.6 | 17.6±0.7 | 16.3±0.7 | 48±4 | 764±250 |

Table 1. Critical dimensions of the hydrogen lithography patterns from the high-resolution STM images (shown in Figure 2), where STM image-broadening artifacts have been corrected. The total pattern areas (in units of squares, or the length-width aspect ratio of the STM-patterned leads) from the source and drain leads between the two inner contact probes (see Figure 1 (a)) are also given. The uncertainties in the number of squares is dominated by the uncertainty in the e-beam alignment between the electrical contacts and the STM-patterned contact pads. The right-most column of the table lists the measured total junction resistances ($R_S + R_D$), where corrections have been taken to eliminate contributions from the source and drain lead sheet resistance. The $R_S + R_D$ for SET-B represents an ohmic resistance where the uncertainty is dominated by uncertainty in estimating the number of squares in the source/drain leads. The $R_S + R_D$ for SET-C to SET-I represents tunneling resistances where the error bars include contributions from both the variation (one standard deviation) in the Coulomb oscillation peak height over the corresponding gating range (-200 mV to 200 mV, see Figure 3 (b)) from multiple gate sweeps and the uncertainty in the subtracted source and drain leads resistance. The uncertainty in the reported dimensions and tunneling resistance values are given as one standard deviation in the distribution of measurement samples.

Figure 2 shows a series of STM images acquired following hydrogen-lithography with surface reconstruction dimer rows clearly visible. Although not all device drain/source electrodes are aligned to the [110] lattice direction, we observe optimal atomic precision lithography by orienting the device in the [110] lattice direction and aligning the geometries of the critical device region (island and tunnel junctions) with the underlying surface reconstructed lattice. For SET-G and SET-I whose tunnel gaps are in the [100] direction, we have corrected for the 45° angle relative to the [110] direction when counting the number of dimer rows in their junction gaps. While attempting to keep lead width and island size identical, we systematically increase the number of dimer row counts within the tunnel junction gap starting from a continuous wire with zero gaps up to SET tunnel gap distances of ~16.2 dimer rows, covering a large range of SET device operation characteristics with respect to tunnel junctions used in QI applications. Because isolated single dangling bonds do not allow dopants to incorporate, we disregard them in quantifying the device geometry. The critical dimensions after STM-imaging correction are summarized in Table 1 for all devices in this study (See Methods for details).

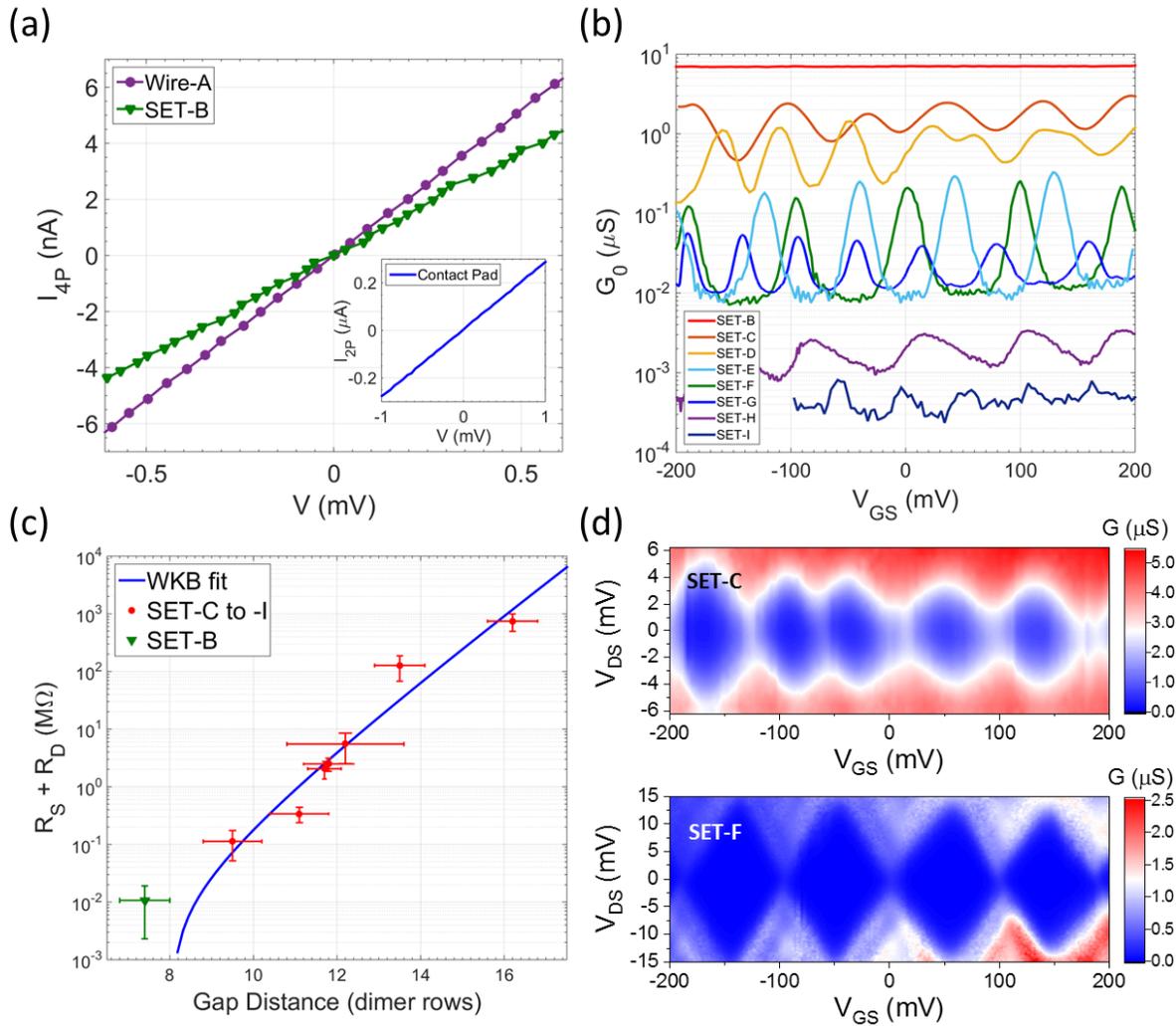

Figure 3. Electrical characterization of the set of devices using a cryostat with a base-temperature of 4 K. (a) Four-point $I_{DS} - V_{DS}$ measurement of Wire-A and SET-B while keeping the gates grounded. Inset: Representative 2-point I-V characteristics (3.5 $k\Omega$) of a device contact pad. (b)

Differential conductance at zero drain-source bias ($G_0$) of the set of SET devices that are measured at T = 4 K. For SET-B to SET-G, $G_0$ is measured using 0.1 mV AC excitation at 11 Hz. For SET-H and SET-I, $G_0$ is numerically estimated from the measured DC charge stability diagrams. From top to bottom: SET-B (red) to SET-I (dark blue). The difference in the oscillation period in gate voltage is due to the variations in gate designs that alter the gate capacitance. (c) The measured total tunneling resistance values $R_S + R_D$ as a function of the lithographically-defined tunnel gap distances. The WKB-fitting is based on the tunneling resistance values from SET-C to SET-I, where the lateral electrical seam width of the electrodes and the rectangular barrier height are taken as free fitting parameters. (d) The measured differential conductance $dI_{DS}/dV_{DS}$ (on a color linear scale) charge stability diagram of SET-C (upper panel) and SET-F (lower panel) at T = 4 K.

In Figure 3 (a), the I-V characteristics of Wire-A exhibit Ohmic behavior with a 4-point resistance of 96.8 $k\Omega$. Considering the actual STM-patterned wire geometry (approximately 57 ± 4 squares between the e-beam patterned voltage contact probes, see Figure 2(a)), this corresponds to a sheet resistance of 1.70 ± 0.15 $k\Omega$ in the STM-patterned electrodes, in excellent agreement with previous results on metallically doped Si:P delta layers.[17] Given the ultrahigh carrier density and small Thomas Fermi screening length[15] in this saturation-doped Si:P system and the relatively large island size[18] of the SETs, we treat the energy spectra in the islands and source and drain leads as continuous ($\Delta E \ll k_B T$, where $\Delta E$ is the energy level separation in the island and source and drain reservoirs) and adopt a metallic description of SET transport.[16] The tunneling rates, $\Gamma_{S,D}$, and the tunneling resistances, $R_{S,D} = \hbar/(2\pi e^2 |A|^2 D_i D_f)$, across the source and drain tunnel barriers can be described using Fermi's golden rule,[19] where $A$ is the tunneling matrix element, $D_{i,f}$ represents the initial and final density of states, $\hbar$ is the reduced Plank's constant, and $e$ is the charge of an electron.

In the following, we show that the total tunneling resistance $R_S + R_D$ of an SET can be extracted by measuring, at zero drain-source DC-bias, the peak amplitudes of the differential conductance Coulomb oscillations, as shown in Figure 3 (b). At $V_{DS} = 0\ V$, the differential conductance Coulomb blockade oscillations reach peaks at $V_{GS} = V_{GS}^{peak} = \left(N + \frac{1}{2}\right)\frac{e}{C_G}$, where $N$ is an integer and $\left(N + \frac{1}{2}\right)e$ represents the effective gating charge when the island Fermi level $\mu_{IS}(N)$ aligns with $\mu_S$ and $\mu_D$. At low temperatures and in the metallic regime, $\Delta E \ll k_B T \ll E_C$, where $E_C = e^2/C_\Sigma$ is the charging energy, and $C_\Sigma = C_S + C_D + C_G$ is the total capacitance (see Table S1 in the supplementary information), and assuming energy independent tunnel rates and density of states in a linear response regime, Beenakker and co-workers [20, 21] have shown that the peak amplitude of the zero-bias differential conductance oscillations in an SET reduces to the following temperature independent expression for arbitrary $R_S$ and $R_D$ values,

$$\left.\frac{dI_{DS}}{dV_{DS}}\right|_{V_{GS}^{peak}} = \frac{e^2\rho}{2}\frac{\Gamma_S\Gamma_D}{\Gamma_S + \Gamma_D} = \frac{1}{2}\frac{G_S G_D}{G_S + G_D} = \frac{1}{2(R_S + R_D)}$$

Equation 1

where $G_S$ and $G_D$ are conductances through the source and the drain tunnel barriers, $\rho$ is the density of state in the metallic island, and the density of states in the leads is embedded in the tunneling rates.

In Figure 3 (b) we observe Coulomb blockade oscillations in all SETs except SET-B. The small gap distance (~7.4 dimer rows ≈ $5.7\ nm$) in SET-B is comparable to twice the Bohr radius, $r \sim 2.5\ nm$, of an isolated P atom in bulk Si,[22] indicating significant wavefunction overlap within the gap regions between the island and the source/drain reservoir. Given that SET-B does not exhibit single electron tunneling behavior (Coulomb oscillations), we estimate the resistance at the junction gaps in this device using 4-point I-V measurement. As shown in Figure 3 (a), SET-B has a linear I-V behavior with the 4-point resistance of $136.7\ k\Omega$. Subtracting the resistance contribution from the source/drain leads (~74 squares) using the estimated sheet resistance (~1.7 $k\Omega$) from Wire-A, we obtain a junction resistance value of ~5.5 ± 4.5 $k\Omega$ per junction in SET-B, which does indeed fall below the resistance quantum (~26 $k\Omega$), and explains the absence of Coulomb blockade behavior. We emphasize that, due to the absence of the Coulomb blockade effect, the estimated resistance at the junctions in SET-B is an ohmic resistance, which should not be confused with the tunneling resistance.

For the rest of the SETs, we extract the total tunneling resistance, $R_S + R_D$, from the Coulomb oscillation peak heights following Equation 1. Figure 3 (c) summarizes the measured junction resistance values (after sheet resistance correction from the source and drain leads) as a function of the averaged gap distances. The tunneling resistance follows a clear exponential relationship with the gap distances. It is notable that a change of only nine dimer rows gives rise to over four orders of magnitude change in the junction resistance. Increasing the gap distance over a small range (from ~7 dimer rows in the gap to ~12) dramatically changes the SET operation from a linear conductance regime (no sign of Coulomb oscillations at ~7 dimer rows separation in SET-B) to a strong tunnel coupling regime (at ~9.5 dimer rows separation in SET-C) to a weak tunnel coupling regime (at ~12 dimer rows separation in SET-F). The relatively strong tunnel coupling in SET-C (see Figure 3(d) upper panel) blurs the charge quantization on the island and introduces finite conductance within the Coulomb diamonds through higher order tunneling processes (co-tunneling).[23] In the weak tunnel coupling regime in SET-F (see Figure 3(d) lower panel), the Coulomb blockade diamonds become very well established. Tuning the tunnel coupling between strong and weak coupling regimes in atomic devices is an essential capability: e.g. for simulating non-local coupling effects in frustrated systems. [24]

It has been found essential for capacitance modeling (See Table S1 in supplementary information) to add a lateral electrical seam [25] and a vertical electrical thickness [26] to the STM-patterned hydrogen-lithography geometry (Figure 2) to account for the Bohr radius and yield the actual "electrical geometry" of the device. We fit the total tunneling resistance ($R_S + R_D$) from SET-B to SET-H as a function of the tunnel gap distance by simulating a single tunnel junction's tunneling resistance (multiplied by two to account for the presence of two junctions) using the WKB method assuming a rectangular barrier shape. [27] (See supplementary information for detailed WKB formulation.) Due to the linear dependence of the WKB tunneling resistance on the tunnel junction cross-sectional area, we ignore the small variations in the STM-patterned junction width, $w$, (see column 3 in Table 1) and adopt an averaged value of $w = 12\ nm$ in the WKB simulation. We account for the "electrical geometry" of the devices by assuming an electrical thickness of $z = 2\ nm$,[26] while treating the lateral electrical seam width, $s$, and the rectangular barrier height, $E_{barr}$, as fitting parameters. We obtain $100 \pm 50\ meV$ as the best-fit barrier height (uncertainty represents two σ), which is in good agreement with the theoretically predicted range of Fermi levels below the Si conduction band edge in highly $\delta$-doped Si:P systems, ~80 meV to ~130 meV, from tight-binding [26] and density functional theory [22] calculations. A similar barrier height value (~80 meV) has also been experimentally determined in a Fowler-Nordheim tunneling regime by Fuhrer's group using a similar STM-patterned Si:P device.[7] We obtain $3.1 \pm 0.4\ nm$ as the best-fit seam width (uncertainty represents two σ), which is in good agreement with the Bohr radius of isolated single phosphorus donors in bulk silicon ($r\sim2.5\ nm$).[22] Using the best-fit seam width from the WKB simulation, we also find good agreement between the experimental and simulated capacitance values from the SETs. (see Table S1 in the supplementary information)

Figure 3 (c) is a key result of this study, clearly demonstrating an exponential scaling of tunneling resistance to the atomic limit. In addition, the data suggests that to obtain tunneling resistance values comparable to those reported in the literature from similar STM-patterned tunnel junctions, [6, 28, 29] we need to pattern our tunnel gaps with smaller gap distances in general. This further emphasizes the improved dopant confinement in the our STM-patterned devices. We highlight that the series of devices shown in Figure 2 were fabricated in series from two different UHV-STM systems. This is important as it further demonstrates atom scale control across similar but non-identical hardware platforms using the same nominal methods and processes.

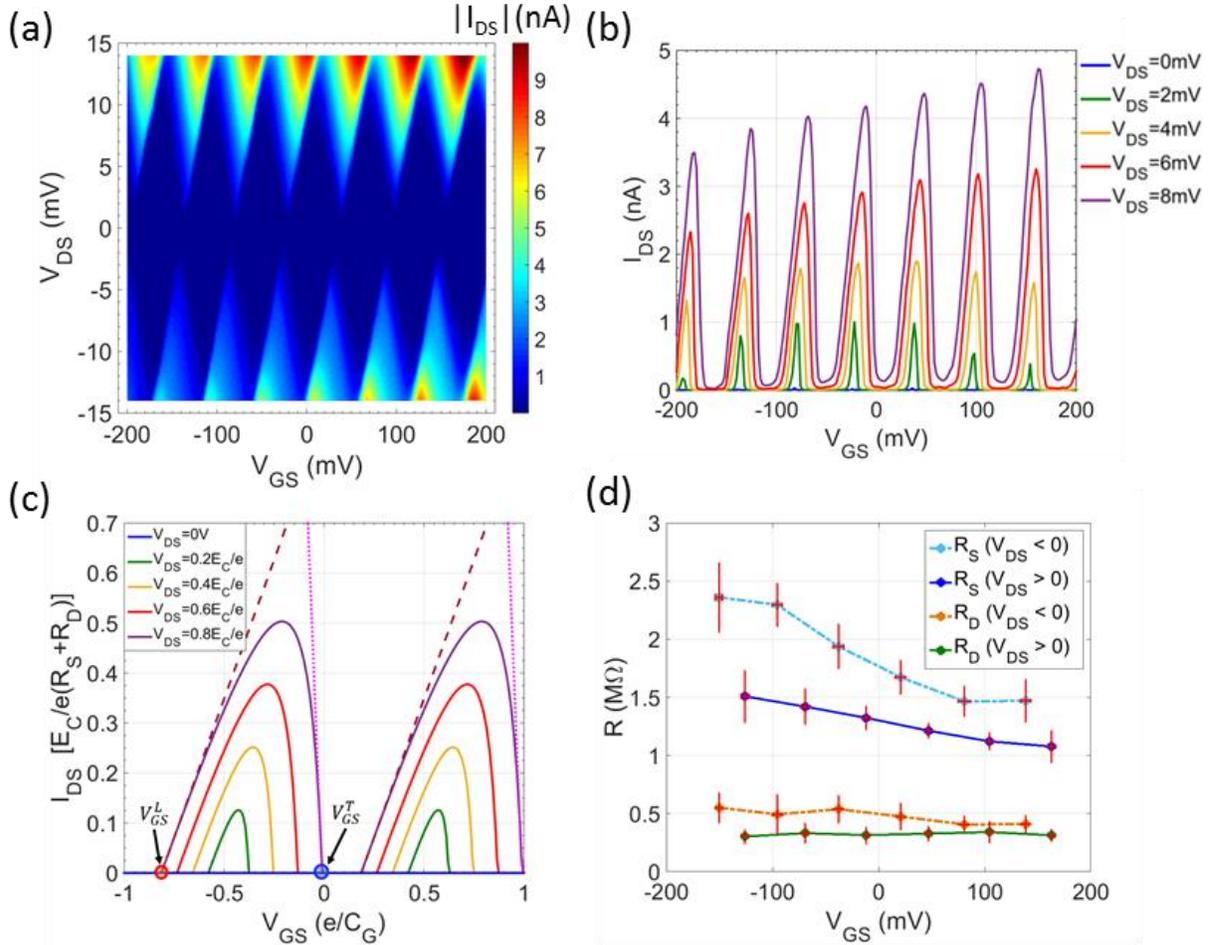

Figure 4. DC measurement of SET-G using a dilution refrigerator with a base-temperature of ~10 mK. (a) The DC-measured charge stability diagram, where the drain-source current $I_{DS}$ is plotted as the absolute values for clarity. (b) The measured Coulomb blockade oscillations at selected drain-source biases. (c) Simulated Coulomb blockade oscillations at positive drain-source bias, assuming asymmetric junction resistances $R_S = 9R_D$. At $V_{DS} = 0.8E_C/e$, the dotted and dashed lines plot the simulated tunneling current through the rate-limiting source and drain tunnel junctions at the leading and trailing edges of the Coulomb oscillation peaks respectively, while ignoring the other junction in series. (d) The extracted junction resistances from Coulomb oscillation peaks along the gate voltage axis. The horizontal and vertical uncertainties at the data points are calculated by averaging the oscillation peak positions and the tunneling resistances at different drain-source biases.

Having demonstrated atom scale control of the tunnel coupling, we now take an additional step to characterize the junction resistance difference in a pair of nominally identical tunnel junctions in SET-G, where both the tunnel gaps have irregular edges and the tunnel gap distances are less well-defined when compared with the tunnel gaps in the other SETs, representing a lower bound of controllability among the SET devices in this study. We present the measured charge stability diagram and finite bias Coulomb oscillations in Figure 4 (a) and (b). In Figure 4 (b), the Coulomb oscillation peaks are asymmetric across the gate voltage. For positive drain-source bias, at the leading edge of the Coulomb oscillation peak of $N \leftrightarrow N+1$ transition, the island spends most of the time unoccupied ($N$). So, the total tunneling rate is limited by tunneling from the source to the island, and thus the total tunneling resistance is dominated by $R_S$. The other three cases are analogous. Figure 4 (c) takes $V_{DS} > 0$ for instance and shows a numerical simulation (at $T = 0\ K$) of $I_{DS}$ vs. $V_{GS}$ at different drain-source bias. The dashed and dotted lines in Figure 4 (c) illustrate the asymptotic slopes at the leading and trailing edges of the Coulomb oscillation peaks at $V_{DS} = 0.8 E_C/e$, which also represent the tunneling current through the rate-limiting source and drain tunnel junctions, respectively, while ignoring the other junction in series. At $T = 0\ K$, the source and drain junction resistances can be derived from the right derivative at the leading edge, where $V_{GS} = V_{GS}^L = \left(N + \frac{1}{2}\right)\frac{e}{C_G} - \frac{C_D}{C_G}V_{DS}$, and from the left derivative at the trailing edge, where $V_{GS} = V_{GS}^T = \left(N + \frac{1}{2}\right)\frac{e}{C_G} + \frac{(C_S+C_G)}{C_G}V_{DS}$, of a Coulomb oscillation peak in $I_{DS}$. This is shown in Equation 2 (see supplementary information for mathematical derivations), again, taking positive drain-source biases for example,

$$\left.\frac{\partial_+ I_{DS}}{\partial V_{GS}}\right|_{V_{GS}^L} = \lim_{\Delta V_{GS} \to 0} \frac{I_{DS}(V_{GS}^L + \Delta V_{GS}) - I_{DS}(V_{GS}^L)}{\Delta V_{GS}} = \frac{C_G}{R_S C_\Sigma}$$

$$\left.\frac{\partial_- I_{DS}}{\partial V_{GS}}\right|_{V_{GS}^T} = \lim_{\Delta V_{GS} \to 0} \frac{I_{DS}(V_{GS}^T) - I_{DS}(V_{GS}^T - \Delta V_{GS})}{\Delta V_{GS}} = -\frac{C_G}{R_D C_\Sigma}$$

Equations 2.

See Table S1 in the supplementary information for the gate and total capacitances, $C_G$ and $C_\Sigma$. To estimate the drain and source tunneling resistances from the Coulomb oscillation peaks that are measured at finite temperatures (Figure 4 (b)), we approximate the asymptotic slopes at the leading and trailing edges by fitting the leading and trailing slopes of the measured Coulomb oscillation peaks and average over a range of $V_{DS}$ bias. (see Figure 4 (d)) We find a factor of approximately four difference in the source and drain tunneling resistances. Possible contributions to this resistance difference include atomic-scale imperfections in the hydrogen lithography of tunnel gaps, the randomness in the dopant incorporation sites within the patterned regions, and unintentional, albeit greatly suppressed, dopant movement at the atomic-scale during encapsulation overgrowth. From the exponential dependence in Figure 3 (c), a factor of four corresponds to an uncertainty in the gap distance of only about half of a dimer row pitch distance, which represents the ultimate spatial resolution (a single atomic site on the Si (100) 2×1 reconstruction surface) and the intrinsic precision limit for the atomically precise hydrogen-lithography.

In summary, we have reported atomic scale control over the tunnel coupling in STM-patterned Si:P single electron transistors. By using the natural surface reconstruction lattice as an atomic ruler, we systematically varied the tunneling gap distances with atomic precision and demonstrated exponential scaling of tunneling resistance to the atomic limit. We emphasize that, critical fabrication steps, such as a defect- and contaminant-free silicon substrate and hydrogen resist formation, atomically abrupt and ultra-clean hydrogen lithography, with dopant incorporation, epitaxial overgrowth, and electrical contact formation that suppress dopant movement at the atomic scale, are all necessary to realize atomic precision devices. This study represents an important step towards fabricating key components needed for high-fidelity silicon quantum circuitry that demands unprecedented precision and reproducibility.

## Methods

The Si:P single electron transistors (SETs) are fabricated on a hydrogen-terminated Si (100) 2×1 substrate ($3 \times 10^{15}/cm^3$ boron doped) in an ultrahigh vacuum (UHV) environment with a base pressure below $4 \times 10^{-9}\ Pascal$ ($3 \times 10^{-11}\ Torr$). Detailed sample preparation, UHV sample cleaning, hydrogen-resist formation, and STM tip fabrication and cleaning procedures have been published elsewhere.[10, 30, 31] A low 1×10$^{-11}$ Torr UHV environment and contamination-free hydrogen-terminated Si surfaces and STM tips are critical to achieving high-stability imaging and hydrogen lithography operation. The device geometry is defined by selectively removing hydrogen resist atoms using an STM tip in the low-bias ($3\sim 5\ V$) and high-current (15~50 nA) regime where the small tip-sample separation allows for a spatially focused tunneling electron beam under the atomic-scale tip apex, creating hydrogen lithographic patterns with atomically abrupt edges. For complete hydrogen desorption within the patterned regions, the typical tip scan velocity and scan-line spacing are 100 nm/sec and 0.5 nm/line respectively. We then saturation-dose the patterned device regions with PH$_3$ followed by a rapid thermal anneal at 350 ℃ for 1 min to incorporate the P dopant atoms into the Si surface lattice sites while preserving the hydrogen resist to confine dopants within the patterned regions. The device is then epitaxially encapsulated with intrinsic Si by using an optimized locking layer process to suppress dopant movement at the atomic-scale during epitaxial overgrowth.[11, 13] The sample is then removed from the UHV system and Ohmic-contacted with e-beam defined palladium silicide contacts.[14] Low-temperature transport measurements are performed using either a closed-cycle cryostat at a base temperature of 4 K or a dilution refrigerator at a base temperature of ~10 mK. For SET-B to SET-G, the zero-DC bias differential conductance ($G_0$) are measured using 0.1 mV AC excitation at 11 Hz. For SET-H and SET-I, $G_0$ is numerically estimated from the measured DC charge stability diagrams. The gate leakage currents are on the order of ~10 pA or less within the gating range used in this study.

We estimate the critical dimensions of the STM-patterned tunnel junctions in a SET from the STM topography images in Figure 2 of the main text, where the gap-distance, $d$, is the average across the full junction width, $w$, using both junctions. The junction width is the average over the island and the first 15 nm of the source and drain leads near the island. The hydrogen lithography and STM-imaging are carried out using different tips and/or under different tip conditions. To eliminate the STM image-broadening due to the convolution between the wavefunctions of the tip apex and Si danging bonds and extract the boundary of the hydrogen-depassivated surface lattice sites, we estimate the image-brodening, $\Delta b$, from the difference between the imaged single dangling bond size, $b$, (full-width at half maximum (FWHM)) and the size of a single dangling bond lattice site, $b_0$, where we have assumed $b_0$ equals half a dimer row pitch. (see Figure 1 (c) in the main text). The image-broadening, $\Delta b = b - b_0$, is then used to correct the critical dimensions that are read out from the half-maximum height positions in the STM topography images.

The theoretical analysis of the transport through SETs is based on an equivalent circuit model (see Figure 1 (d)) under a constant interaction approximation. The analytical expressions regarding the equilibrium drain-source conductance in the main text are derived using the standard Orthodox theory under a two-state approximation.[18, 32]

## Supplementary Information

**Modeling the Tunnel Barriers Using the WKB Method**

We fit the measured tunneling resistance $R_S + R_D$ as a function of the STM-patterned tunnel gap distance, $d$, using the well-known Simmons' WKB formulation in the low-bias (linear response) regime. [27, 33] We adopt an ideal rectangular barrier shape ignoring the image force correction to the barrier potential when an electron approaches the dielectric barrier interface. The barrier height, $E_{barr}$, is defined as the energy difference between the electrode reservoirs' Fermi level and the conduction band minimum. We expect exponential dependence of the tunnel conductance on both the barrier height and barrier width, whereas a linear dependence on the tunneling cross-section area is expected. Therefore, the slight width variation among the fabricated tunneling junctions is assumed to have minor effects on the tunnel conductance. We assume a uniform electrical thickness $z = 2\ nm$ for the STM-patterned device components. To account for the finite electron density extension beyond the hydrogen-lithography patterns in the lateral directions, we add a uniform lateral seam, $s$, to the device pattern. We adopt an averaged width of $w = 12\ nm$ as the STM-patterned junction width. Therefore, the electrical junction width and the electrical junction gap distance are expressed as $(w + 2s)$ and $(d - 2s)$ respectively. $(w + 2s)z$ represents the electrical tunnel junction cross-sectional area. The lateral seam width, $s$, and the rectangular tunnel barrier height, $E_{barr}$, are treated as fitting parameters. The WKB tunneling resistance, $R_T$, in the low-bias regime is expressed in Equation S1.[27, 33]

$$\frac{1}{R_T} = \frac{[(w + 2s)z]\sqrt{2m^*E_{barr}}}{(d - 2s)} \left(e/h\right)^2 \exp\left[-\frac{4\pi(d - 2s)}{h}\sqrt{2m^*E_{barr}}\right]$$

Equation S1.

Where $h$ is Plank's constant, $e$ is the charge of a single electron, and $m^*$ is the effective mass of the conducting electrons. Conductivity in the degenerately $\delta$-doped Si:P electrodes is assumed to be dominated by the lowest energy sub-bands, with effective mass $m^* = 0.21m_e$ as measured by Miwa *et al.* using direct spectroscopic measurement in blanket $\delta$-doped Si:P layers,[34] where $m_e$ is the free electron mass. We point out that, at a given barrier height $E_{barr}$, the dependence of WKB tunneling resistance, $R_T$, on the gap distance, $d$, deviates from an ideal exponential behavior, especially at small gap distances, due to the pre-factor in front of the exponential term in Equation S1.

**Comparison between the Measured and Simulated Capacitances in STM-patterned SET Devices**

Capacitance modeling of STM-patterned Si:P devices has demonstrated success in accurately predicting the device electrostatics down to the atomic scale.[25] Table S1 compares the experimentally observed SET capacitances and the simulated capacitances, where the device components are treated as metallic sheets in the shape of the "electrical geometry" of the device.[25, 35] A uniform electrical thickness of z=2 nm in the z-direction is assumed for both the Simulation 1 and Simulation 2. No lateral electrical seam is added to the hydrogen lithography pattern in Simulation 1. The simulated capacitances from Simulation 1 agree poorly with the measured capacitances. In Simulation 2, a lateral electrical seam width of 3.1 nm from the WKB tunneling resistance fit is added to the STM-patterned device geometry, which significantly improves the agreement between the simulated and measured capacitances.

|  | $E_C$ (meV) | $C_\Sigma$ (aF) | $C_G$ (aF) | $C_S$ (aF) | $C_D$ (aF) |
|---|---|---|---|---|---|
| Experiment | $11.9 \pm 0.3$ | $13.5 \pm 0.3$ | $2.8 \pm 0.2$ | $5.0 \pm 0.3$ | $5.7 \pm 0.3$ |
| Simulation 1 (no seam) | 19.5 | 8.2 | 2.6 | 2.8 | 2.8 |
| Simulation 2 (with 3.1 nm seam) | 10.5 | 15.3 | 3.2 | 6.0 | 6.1 |

Table S1. The experimental and simulated charging energy and capacitances of SET-G. The experimental capacitances are extracted using the height and width of the measured Coulomb diamonds (Figure 4 (a)) as well as the slopes of the positive and negative diamond edges.[16] The uncertainties result from the experimental determination of the Coulomb diamond dimensions from the measured charge stability diagrams while extracting the experimental capacitances. The capacitance simulation is carried out using a finite-element 3D Poisson solver, FastCap.[36,37]

**Quantifying Individual Junction Resistances in a Metallic SET**

Following the well-established Orthodox theory for a metallic SET,[19] the tunneling resistance across the individual tunnel barriers can be extracted from the peak shapes of Coulomb oscillations in $I_{DS}$. In this section, we derive the explicit expressions in Equation 2 of the main text using an analytical model that was first proposed by Inokawa and Takahashi.[32]

The tunneling probability through an SET is determined by the change in the SET's Helmholtz's free energy $F = U - W$, where $U$ is the total electrostatic energy stored in the system and $W$ is the work done by voltage sources, due to a single electron tunneling event. Following the constant interaction model in a metallic regime (See Figure 1 (d) in the main text), the change in $F$ when an electron tunnels from the source/drain electrodes to the island and transitions the number of excess electrons on the island from $N$ to $N + 1$ can be expressed as $\Delta F_{S,D}^{N+1,N} = -\mu_{S,D} + \mu_{IS}(N)$, where $\mu_{S,D}$ and $\mu_{IS}(N)$ are the chemical potential of the source/drain leads and an SET island with $N$ excess electrons.[38]

In the zero-temperature limit, $T = 0\ K$, the tunneling rates can be expressed using Fermi's golden rule.

$$\Gamma_{S,D}^{N+1,N} = \frac{1}{R_{S,D} e^2} (-\Delta F_{S,D}^{N+1,N}) \Theta(\Delta F_{S,D}^{N+1,N})$$

$$\Gamma_{S,D}^{N,N+1} = \frac{1}{R_{S,D} e^2} (-\Delta F_{S,D}^{N,N+1}) \Theta(\Delta F_{S,D}^{N,N+1})$$

Equation S2

Where $\Theta(x)$ is a unit step function. For simplicity, we have assumed the single electron tunneling events to be elastic without electromagnetic interactions between the tunneling electron and the environmental impedance.[39]

In an equilibrium condition, the stationary occupancy probability, $P(N)$, of the SET island (with $N$ excess electrons) can be derived by requiring $dP(N)/dt = 0$ in a steady state master equation [16] and obtaining $P(N)\left(\Gamma_S^{N+1,N} + \Gamma_D^{N+1,N}\right) = P(N+1)\left(\Gamma_S^{N,N+1} + \Gamma_D^{N,N+1}\right)$. At low-temperatures where $k_B T \ll E_C$, only the two most-probable charge states dominate the SET island occupancy at a given bias. Adopting a two-state approximation,[32] $P(N) + P(N+1) = 1$, an analytical expression of the total drain-source current through the SET can be obtained,

$$I_{DS}(N) = -eP(N)\Gamma_D^{N+1,N} + eP(N+1)\Gamma_D^{N,N+1}$$

$$= e \frac{\Gamma_D^{N,N+1}\Gamma_S^{N+1,N} - \Gamma_D^{N+1,N}\Gamma_S^{N,N+1}}{\Gamma_D^{N+1,N} + \Gamma_S^{N+1,N} + \Gamma_D^{N,N+1} + \Gamma_S^{N,N+1}}$$

Equation S3

Using the expression of $\mu_{IS}(N)$ $from$ the constant interaction model,[38] we have,

$$I_{DS}|_{T=0}$$
$$= \frac{1}{C_\Sigma} \frac{\left[\frac{e}{2} + (Ne - Q_0) + (C_S + C_G)V_{DS} - C_G V_{GS}\right]\left[\frac{e}{2} + (Ne - Q_0) - C_D V_{DS} - C_G V_{GS}\right]}{R_D \left[\frac{e}{2} + (Ne - Q_0) - C_D V_{DS} - C_G V_{GS}\right] - R_S \left[\frac{e}{2} + (Ne - Q_0) + (C_S + C_G)V_{DS} - C_G V_{GS}\right]}$$

Equation S4

where $Q_0$ ($|Q_0| \leq \frac{e}{2}$) represents a fractional electron charge that is present on the island when the voltage electrodes are floating, typically due to background charges from the environment.

Taking $V_{DS} > 0$ at $T = 0\,K$ for instance, the source and drain junction tunneling resistances can be derived from the right derivative at the leading edge, where $V_{GS} = V_{GS}^L = \left(N + \frac{1}{2}\right)\frac{e}{C_G} - \frac{C_D}{C_G}V_{DS}$, and from the left derivative at the trailing edge, where $V_{GS} = V_{GS}^T = \left(N + \frac{1}{2}\right)\frac{e}{C_G} + \frac{(C_S+C_G)}{C_G}V_{DS}$, of a Coulomb oscillation peak in $I_{DS}(V_{GS})$. According to Equation S4 (assuming $Q_0 = 0$), the right derivative at the leading edge, where $V_{GS} = V_{GS}^L$, has the following expression,

$$\left.\frac{\partial_+ I_{DS}}{\partial V_{GS}}\right|_{V_{GS}^L} = \lim_{\Delta V_{GS} \to 0} \frac{I_{DS}(V_{GS}^L + \Delta V_{GS}) - I_{DS}(V_{GS}^L)}{\Delta V_{GS}}$$

$$= \lim_{\Delta V_{GS} \to 0} \frac{1}{C_\Sigma \Delta V_{GS}} \frac{(C_\Sigma V_{DS} - C_G \Delta V_{GS})(-C_G \Delta V_{GS})}{R_D(-C_G \Delta V_{GS}) - R_S(C_\Sigma V_{DS} - C_G \Delta V_{GS})} = \frac{C_G}{R_S C_\Sigma}$$

Equation S5

Similarly, the left derivative at the trailing edge, where $V_{GS} = V_{GS}^T$, has the following expression,

$$\left.\frac{\partial_- I_{DS}}{\partial V_{GS}}\right|_{V_{GS}^T} = \lim_{\Delta V_{GS} \to 0} \frac{I_{DS}(V_{GS}^T) - I_{DS}(V_{GS}^T - \Delta V_{GS})}{\Delta V_{GS}}$$

$$= \lim_{\Delta V_{GS} \to 0} \frac{-1}{C_\Sigma \Delta V_{GS}} \frac{(C_G \Delta V_{GS})(C_G \Delta V_{GS} - C_\Sigma V_{DS})}{R_D(C_G \Delta V_{GS} - C_\Sigma V_{DS}) - R_S(C_G \Delta V_{GS})} = \frac{-C_G}{R_D C_\Sigma}$$

Equation S6

To estimate the drain and source tunneling resistances from the Coulomb oscillation peaks that are measured at finite temperatures, we approximate the asymptotic slopes at the leading and trailing edges by fitting the leading and trailing slopes of the measured Coulomb oscillation peaks.

## Conflicts of Interest

There are no conflicts of interest to declare.

## Acknowledgements

This work was sponsored by the Innovations in Measurement Science (IMS) program at NIST: Atom-based Devices: single atom transistors to solid state quantum computing. This work is also funded in part by the Department of Energy. We thank Daniel Walkup and John Kramar for valuable comments and discussions. This work was performed in part at the Center for Nanoscale Science and Technology NanoFab at NIST.